# Splitting of temperature distributions due to dual-channel photon heat exchange in many-body systems

Ming-Jian He[1,2], Xue Guo[1,2], Hong Qi[1,2,*], Ivan Latella[3,**], and He-Ping Tan[1,2]

1 School of Energy Science and Engineering, Harbin Institute of Technology, Harbin 150001, P. R. China

2 Key Laboratory of Aerospace Thermophysics, Ministry of Industry and Information Technology, Harbin 150001, P. R. China

3 Department of Condensed Matter Physics, University of Barcelona, Martí i Franquès 1, 08028 Barcelona, Spain

*Corresponding authors: Email: qihong@hit.edu.cn (H. Qi), ilatella@ub.edu (I. Latella)

**Abstract:** We investigate the radiative heat transfer and spatial distributions of stationary temperatures in periodic many-body systems composed of alternating slabs of two different materials. We show that temperature distributions exhibit an alternating spatial pattern and split into two distinct components, with each component corresponding to one of the two materials. Spatial temperature variations following the periodicity of the structure can be attributed to a dual-channel photon heat exchange through a long-range coupling of electromagnetic modes supported by bodies of the same material. We also analyze the thermal relaxation of the temperatures in the system to verify potential applications in dynamical situations. The results reveal that tunable nonmonotonic temperature variations can be also designed and utilized at a transient mode. The dual-channel mechanism to control temperature distributions proposed in the present work may pave new avenues for prospective applications in nano devices, especially for thermal photon-driven logic circuitry and thermal management.

# I. INTRODUCTION

The radiative heat flux can exceed the limit defined by the Planck's law when the transport distance is much smaller or comparable to the peak value of the characteristic wavelength [1-4]. In this regime, the thermal radiation energy exchange between objects is referred to as near-field radiative heat transfer (NFRHT) [5-8]. Over the past few decades, with the development of nanoscience and the gradual maturity of nanotechnology, the theory of NFRHT has been experimentally confirmed [1, 9], which has greatly stimulated the interest of the scholars in this field. In recent years, a large amount of research was carried out leading to new theoretical results [10-15] and experimental works [1, 9, 16], most of which have focused on the radiative heat transfer in systems composed of two interacting bodies [17-24].

While the study of two-body systems has been extensive, it was only in recent years that research on NFRHT in many-body systems began to emerge and expand [25-34]. The NFRHT in these systems involves cooperative effects between the emitters which lead to emergent phenomena that are absent in the two-body case. Thus, describing the radiative heat transfer in many-body structures requires a theoretical framework that properly incorporates the multibody thermal emission [25]. The consequences of many-body radiative interactions can be observed even in relatively simple systems consisting of three bodies [35], including the amplification or suppression of heat transfer [36, 37], as well as thermalization effects at small separations [38]. Compared with two-body systems, the features of three-body systems have resulted in a wider range of novel physical phenomena, providing diverse application scenarios. Taking advantage of thermal photon tunneling in near-field configurations, functional devices based on three-body systems have been proposed [39-42].

As the number of thermal emitters increases, additional degrees of freedom become increasingly important and enable the consideration of more complex situations. For systems with an arbitrary number of bodies and planar geometry, the NFRHT can be described by means of a Landauer-like formalism which is based on the scattering theory and the fluctuational electrodynamics approach [43]. Within this framework, a transition from superdiffusive to ballistic transport was demonstrated in dense many-body systems [44], the stationary temperature distribution in a multilayered graphene system was investigated and temperature steps near the boundaries of the system were predicted [45], and the concept of thermal barrier was proposed by combining different materials in many-body structures [46]. These works emphasize the fact that it is possible to design many-body systems to achieve desired functionalities and thermal properties by suitably setting the distribution of materials in the structure.

In planar many-body systems with homogeneous properties subject to an imposed temperature difference at the system boundaries, a monotonic temperature distribution is always expected in the steady state as long as there are no internal heat sources or sinks. A monotonic distribution can also be expected in heterogeneous systems, without internal heat sources or sinks, if the heat exchange through the bodies is due to first-neighbor interactions only. In this case, the stationary temperature of a given object must necessarily lie between the temperatures of adjacent bodies. However, if heat exchange in heterogeneous systems proceeds through a long-range coupling between its constituents, that is, beyond first neighbors, nonmonotonic temperature distributions may be realized due to efficient coupling between distant objects whose properties are different from those in the close vicinity. In the present work this behavior is achieved by considering two materials supporting surface phonon polaritons (SPhPs) at different frequencies, implemented in a periodic structure with slabs of finite thickness of alternating materials. The finite nature of the slabs allows for transmission of the electromagnetic field, thereby permitting a long-range coupling between objects made of the same material. This coupling introduces two channels for radiative heat transport, where photons are primarily guided by the coupling between objects of the same material. As a result, these effective channels enable efficient radiative heat transfer between such objects, facilitating the exchange of thermal energy over long distances. Our findings reveal that this dual-channel heat transport induces a splitting of the temperature distribution into two distinct components, with each component corresponding to one of the two materials. Consequently, overall spatial distributions of stationary temperatures display an alternating pattern. This phenomenon can be utilized to improve our ability to modulate heat and control the temperature at the nanoscale, offering new possibilities for radiative thermal management.

The paper is structured as follows. In Section II, we introduce the physical system and the formalism to describe the radiative heat transfer. In Section III, temperature profiles in many-body system are investigated considering different material configurations. To explore the underlying mechanism in the splitting of temperature distributions, the dual-channel photon heat exchange is analyzed in detail. Furthermore, we test the time relaxation and transient temperature states of the proposed system, comparing the results with those of a single-material system. Finally, our conclusions are presented in Section IV.

## II. PHYSICAL SYSTEM AND MODEL

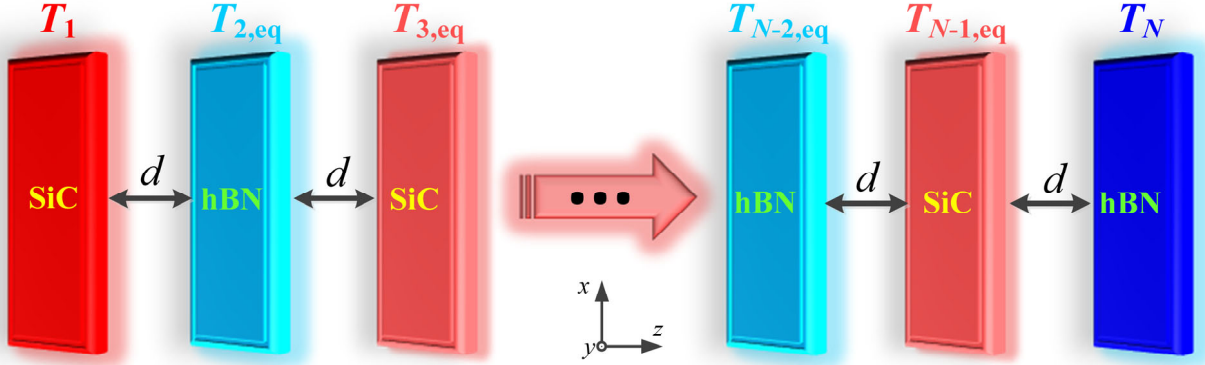

Fig. 1. Schematic of a many-body system composed of $N$ planar slabs, and with two external slabs at fixed temperatures $T_1 = 400$ K (red) and $T_N = 300$ K (blue). The slabs are made of two different materials, SiC and hBN, which are alternately arranged. The system is assumed to be placed in vacuum and the all the separation distances between adjacent slabs $d$ are identical. In the steady state, all the internal slabs reach a local equilibrium temperatures $T_{j,\,\mathrm{eq}}$ ($1 < j < N$).

Here we consider a many-body system composed of $N$ planar slabs, orthogonal to the $z$ axis located at positions $z_j$ and assumed to be infinite in the $x$ and $y$ directions, as illustrated in Fig. 1. The system is assumed to be placed in vacuum and all the internal separation distances $d$ are identical. For the sake of illustration, the thicknesses $\delta$ of the slabs are assumed to be equal, which is taken as $\delta = 200$ nm in the following results. The external slabs, i.e., bodies 1 and $N$, are held at constant temperatures $T_1 = 400$ K and $T_N = 300$ K, respectively. We assume that the system is thermalized in an environment at temperature $T_B = T_N = 300$ K. In the stationary state, the internal slabs are allowed to reach their own equilibrium temperature $T_{j,\,\mathrm{eq}}$ ($1 < j < N$) which is assumed to be uniform within the bodies. Here, the slabs are made of two different polar materials, silicon carbide (SiC) and hexagonal boron nitride (hBN), which are alternately arranged. The permittivity of SiC and hBN at frequency $\omega$ can be described by the Drude-Lorentz model

$$\varepsilon(\omega) = \varepsilon_\infty \frac{\omega_\mathrm{L}^2 - \omega^2 - i\Gamma\omega}{\omega_\mathrm{T}^2 - \omega^2 - i\Gamma\omega} \tag{1}$$

where $\varepsilon_\infty$ is the high frequency dielectric constant, $\omega_\mathrm{L}$ is the longitudinal optical phonon frequency, $\omega_\mathrm{T}$ is the transverse optical phonon frequency, and $\Gamma$ is the damping rate. The parameters are taken as $\varepsilon_\infty = 6.7$, $\omega_L = 1.83 \times 10^{14}$ rad/s, $\omega_T = 1.49 \times 10^{14}$ rad/s, $\Gamma = 8.97 \times 10^{11}$ rad/s for SiC, and $\varepsilon_\infty = 4.9$, $\omega_L = 3.03 \times 10^{14}$ rad/s, $\omega_T = 2.57 \times 10^{14}$ rad/s, $\Gamma = 1 \times 10^{12}$ rad/s for hBN, respectively.

Radiative heat transport in the system arises due to the thermal bias imposed on the first and last bodies, which can be quantified by considering the averaged component of the Poynting vector in the $z$ direction. The net radiative flux received by body $j$ can be written as a sum over the energy exchanged with every other body $\varphi_{l,j}$ [43],

$$\varphi_j = \sum_{l \neq j} \varphi_{l,j} = \frac{1}{4\pi^2} \sum_{l \neq j} \int_0^\infty \hbar \omega n_{l,j} d\omega \int_0^\infty \kappa \xi^{l,j}(\omega, \kappa) d\kappa \tag{2}$$

where $l \neq j$ runs from 1 to $N$, $\hbar$ is the reduced Planck constant and $n_{l,j}(\omega) = n_l(\omega) - n_j(\omega)$ denotes the difference between the two mean photon occupation numbers $n_{l/j}(\omega) = \left[\exp(\hbar\omega / k_B T_{l/j}) - 1\right]^{-1}$, $k_B$ being the Boltzmann constant. The computation of $\xi^{l,j}(\omega, \kappa)$ contains the sum of the p and s polarizations. Here $\kappa$ is the component of the wave vector parallel to the interfaces and $\xi^{l,j}(\omega, \kappa)$ describes the energy transmission coefficients between body $l$ and $j$. The expression of $\xi^{l,j}(\omega, \kappa)$ is obtained from the many-body theory, which is based on the combination of the scattering theory and the fluctuational-electrodynamics approach in many-body systems. First, the single reflection and transmission coefficients of body $j$ can be written as

$$\rho_j = \frac{r_j \left(1 - e^{2ik_{zj}\delta_j}\right)}{1 - r_j^2 e^{2ik_{zj}\delta_j}} \tag{3}$$

$$\tau_j = \frac{\left(1 - r_j^2\right) e^{ik_{zj}\delta_j}}{1 - r_j^2 e^{2ik_{zj}\delta_j}} \tag{4}$$

where $k_{zj}$ is the $z$ component of the wave vector inside body $j$ and $r_j$ is the vacuum-medium Fresnel reflection coefficients of body $j$. Here we consider consecutive planar slabs having indexes from $j$ to $m$ (with $j$ smaller than $m$). The many-body scattering reflection and transmission coefficients for them, representing the analogues of $\rho_j$ and $\tau_j$ for a single body, are given as

$$\begin{aligned} \rho_+^{j \to m} &= \hat{\rho}_+^{j \to m} e^{-ik_z(\delta_j + 2z_m)} \\ \rho_-^{j \to m} &= \hat{\rho}_-^{j \to m} e^{-ik_z(\delta_j - 2z_j)} \\ \tau^{j \to m} &= \hat{\tau}^{j \to m} \exp\left[-(m - j + 1)ik_z \delta_j\right] \end{aligned} \tag{5}$$

where

$$\hat{\rho}_+^{j \to m} = \rho_m + (\tau_m)^2 \hat{\rho}_+^{j \to m-1} u^{j \to m-1,m} e^{2ik_z d}$$
$$\hat{\rho}_-^{j \to m} = \rho_j + (\tau_j)^2 \hat{\rho}_-^{j+1 \to m} u^{j,j+1 \to m} e^{2ik_z d} \qquad (6)$$
$$\hat{\tau}^{j \to m} = \hat{\tau}^{j \to m-1} u^{j \to m-1,m} \tau_m$$

and

$$u^{j \to m-1,m} = \left(1 - \hat{\rho}_+^{j \to m-1} \rho_m e^{2ik_z d}\right)^{-1}$$
$$u^{j,j+1 \to m} = \left(1 - \rho_j \hat{\rho}_-^{j+1 \to m} e^{2ik_z d}\right)^{-1} \qquad (7)$$

The subscripts "+" and "-" in these expressions denote the direction of propagation or decay of the outgoing field.

To obtain the many-body scattering coefficients given above, one needs to perform an iterative calculation and take $\hat{\rho}_+^j = \hat{\rho}_-^j = \rho_j, \hat{\tau}^j = \tau^j$ for a single body. Then, the Landauer-like energy transmission coefficients between bodies $l$ and $j$ can be subsequently computed based on the obtained many body scattering coefficients,

$$\xi^{l,j} = \hat{\xi}_{j-1}^l - \hat{\xi}_{j-1}^{l-1} - \hat{\xi}_j^l + \hat{\xi}_j^{l-1} \qquad (8)$$

where [43]

$$\hat{\xi}_\gamma^j = \frac{4\left|\tau^{j+1 \to \gamma}\right|^2 \operatorname{Im}\left(\rho_+^{0 \to j}\right) \operatorname{Im}\left(\rho_-^{\gamma+1 \to N}\right)}{\left|1 - \rho_+^{0 \to \gamma} \rho_-^{\gamma+1 \to N}\right|^2 \left|1 - \rho_+^{0 \to j} \rho_-^{j+1 \to \gamma}\right|^2}, j < \gamma$$

$$\hat{\xi}_\gamma^\gamma = \frac{4 \operatorname{Im}\left(\rho_+^{0 \to \gamma}\right) \operatorname{Im}\left(\rho_-^{\gamma+1 \to N}\right)}{\left|1 - \rho_+^{0 \to \gamma} \rho_-^{\gamma+1 \to N}\right|^2} \qquad (9)$$

$$\hat{\xi}_\gamma^j = \frac{4\left|\tau^{\gamma+1 \to j}\right|^2 \operatorname{Im}\left(\rho_+^{0 \to \gamma}\right) \operatorname{Im}\left(\rho_-^{j+1 \to N}\right)}{\left|1 - \rho_+^{0 \to j} \rho_-^{j+1 \to N}\right|^2 \left|1 - \rho_+^{0 \to \gamma} \rho_-^{\gamma+1 \to j}\right|^2}, j > \gamma$$

## III. RESULTS AND DISCUSSION

Considering the system and the theoretical approach described in the preceding section, here we analyze in detail the spatial distribution of temperatures in the system as well as the main features of the NFRHT in situations of interest. To complement our results, at the end of this section we also study the time relaxation of the temperatures in relevant scenarios and describe transient states.

### A. SPLITTING OF TEMPERATURE DISTRIBUTIONS

To demonstrate the splitting of the temperature distributions, we plot the equilibrium temperature profiles in Fig.2 for many-body systems in two different configurations: systems with all slabs made of the same material and a system with alternating slabs of different materials. The system reaches its steady state when the net flux received by each internal slab vanishes, that is, all the internal slabs reach their own equilibrium (stationary) temperature $T_{eq,j}$ ($j=2, \ldots, N-1$) [47]. The local equilibrium temperatures $T_{eq,j}$ of the internal layers can be obtained by means of an iterative procedure [43]. On the one hand, Figs. 2(a) and (b) denote the equilibrium temperature distribution of a many-body system with slabs made of SiC and for a system with slabs of hBN, respectively. On the other hand, Fig. 2(c) illustrates the equilibrium temperature distributions of a many-body system comprising alternating slabs of SiC and hBN. In these figures, the profiles in different colors are the temperature results for different separation distances between adjacent slabs. The horizontal axis in these figures denotes the positions of different slabs in the system, where $z_j/z_N = 0$ specifies the left external slab and $z_j/z_N = 1$ denotes the position of the right external slab, which are numbered with indexes 1 and $N$ in Fig.1, respectively.

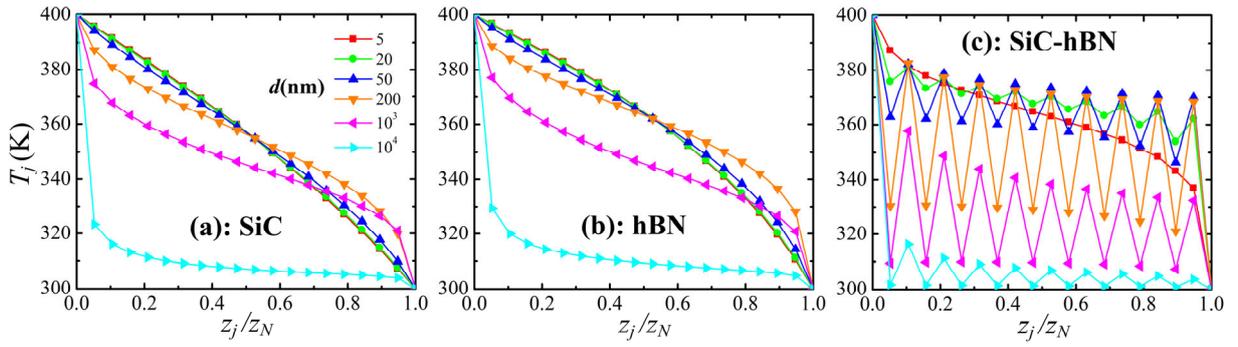

Fig. 2. Equilibrium temperature profiles for a many-body systems composed of $N = 20$ planar bodies: (a) slabs made of SiC; (b) slabs made of hBN; (c) system with alternating slabs of SiC and hBN, i.e., the number of SiC and hBN slabs are both $N/2$.

As can be seen from the results in Figs. 2(a) and (b) for many-body systems made with all slabs made of the same material (either SiC or hBN), the temperature profiles vary with the separation distance and are nearly the same when $d < 200$ nm. As the separation distance increases, the heat transfer between the slabs decay, approaching the far-field regime at large separations. The equilibrium temperatures of the slabs greatly decrease at large separations and get close to the temperature of the environment which is the same as that of the cold slab, $T_N$ = 300 K. Either for SiC or hBN, the system made of uniform material reveals a monotonic trend in the equilibrium temperature distribution.

Let us now discuss the results for the system made of alternating slabs of SiC and hBN in Fig. 2(c). The profile for $d$ = 5 nm is monotonic in the positions of the slabs, like those shown in Figs. 2(a) and (b). As explained more in detail in the next section, this is due to the fact that first-neighbor interactions are strong enough to shape the temperature distribution at very short separations, as bodies tend to thermalize [38]. As clearly evidenced in Fig. 2(c), the temperature profiles exhibit an alternating spatial pattern modulated by the global temperature decay when the separation distance is $d$ > 5 nm. As $d$ gets larger, the alternating pattern in the temperatures becomes increasingly apparent and pronounced. Remarkably, the alternating pattern results from the composition of the temperature profiles associated with the SiC and hBN slabs separately. In other words, the overall temperature distribution splits into two different components that correspond to the two different materials. Furthermore, as the separation is further increased, all the equilibrium temperatures decrease because the heat transfer decays when approaching the far-field regime and the interaction with the environmental bath becomes appreciable. When $d$ is 10 μm, the temperatures of the SiC slabs get much smaller than the left external SiC slab, while the temperatures of the hBN slabs notably decrease and get nearly the same as that of the right external hBN slab with $T_N$ = 300 K (the same temperature as the environment). The above temperature profiles for different separation distances demonstrate that, by tuning the density of the many-body system, different magnitudes of the temperature profiles can be obtained which are characterized by distinctive splitting amplitudes.

## B. DUAL-CHANNEL PHOTON HEAT EXCHANGE

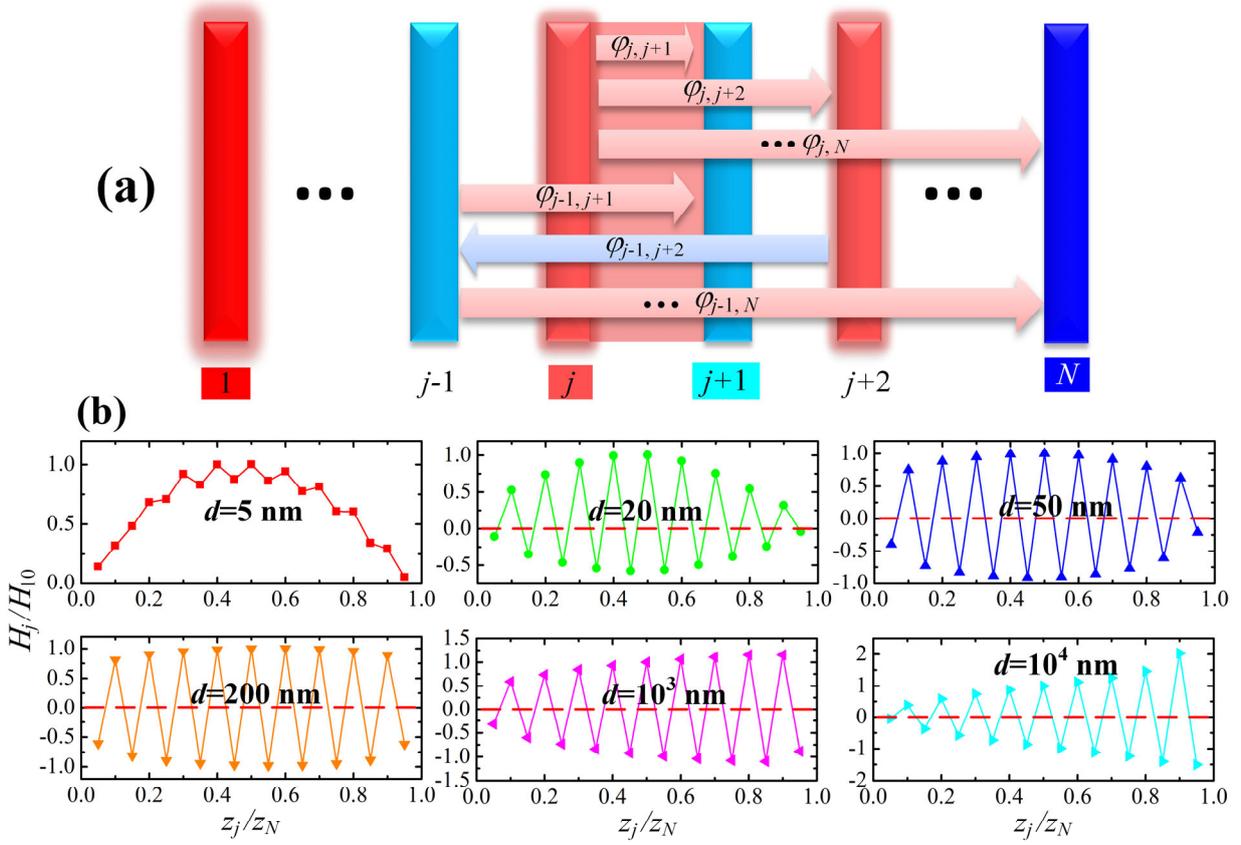

Fig. 3. (a) Schematic of the different contributions $\varphi_{m,n}$ to the radiative flux through the gap between slab $j$ and $j$+1. In (b), we show the ratio of the local heat transfer coefficient at different vacuum gaps to that at the middle position, which is between slabs numbered with 10 and 11 for an $N = 20$ system.

To explore thermal properties of the structure and to quantitatively evaluate the heat transport inside the system, we define a local heat transfer coefficient (LHTC) given by

$$H_j = \frac{\Phi_{j,j+1}}{\Delta T_{j,j+1}}, \qquad (10)$$

where $\Phi_{j,j+1}$ denotes the net radiative flux through the gap between slab $j$ and $j$+1, and $\Delta T_{j,\,j+1}$ denotes the temperature difference between body $j$ and $j$+1. The net radiative flux can be written as a sum of the radiative heat flux transferred from slabs 1-$j$ to $j$+1-$N$, i.e., $\Phi_{j,j+1} = \sum_{m=1}^{j} \sum_{n=j+1}^{N} \varphi_{m,n}$, schematically illustrated in Fig. 3(a). Here $\varphi_{m,n}$ denotes the radiative heat flux exchanged between bodies $m$ and $n$, which can positive or negative depending on the temperature difference between the two bodies. For example, in Fig. 3(a), we schematically plot arrows to show the radiative heat flux exchanged between different bodies considering that the temperature profile is nonmonotonic, realizing the alternating pattern. Bodies $j$-1 and $j$+1 are colored blue in this figure, which means they have a relatively lower equilibrium temperature, just like the hBN slabs in the situations represented in Fig. 2(c). Bodies $j$ and $j$+2 in Fig. 3(a) are colored red and have a relatively higher equilibrium temperature, like the SiC slabs. Based on that, the directions of the arrows can be viewed as the direction of the net heat flux exchanged between different bodies. Considering the nonmonotonic temperature profiles in Fig. 2(c), for instance, $\varphi_{j-1, j+2}$ can be negative in some cases and the associated arrow should be reversed. In contrast, note that the net flux $\Phi_{j,j+1}$ in any of the vacuum gaps always proceeds form left to right in our setup (it is positive), since the temperature of the left external body is higher than that of the right external body.

The LHTC $H_j$ can be viewed as a measure of the thermal freedom to transport heat at different positions in the system. According to the definition in Eq. (10), this coefficient can be negative if the sign of the local temperature difference $\Delta T_{j, j+1}$ is different from the sign of the global temperature difference imposed at the boundaries of the system. In Fig. 3(b), we plot a ratio of LHTC at different vacuum gaps to that at the middle position, which is between slabs numbered with 10 and 11 for an $N$ = 20 system. For $d$ = 5 nm, the ratio of LHTC is positive at different positions inside the system, which is consistent with the monotonic temperature profile shown in Fig. 2(c). This ratio is relatively smaller at the two extremes of the system and takes larger values at the middle positions. This means that the local thermal resistance at the middle is smaller than that near the two external parts of the system. For $d$ > 5 nm, the ratio of LHTC shows a completely different varying behavior due to the splitting of the temperature distribution. Moreover, the profiles for $d$ = 20 nm, 50 nm, and 200 nm are nearly symmetric with the positions. The positive and negative points show a similar trend, and their absolute values are also lower at two sides and higher at the middle of the system. However, for $d$ = $10^3$ nm and $10^4$ nm, this symmetry gets distorted due to the different behavior of the heat transfer in near-field and far-field regimes and to the influence of the environmental bath. The behavior of the LHTC and the local transport properties are governed by effective channels for heat exchange in the system associated with material properties, as we discuss next.

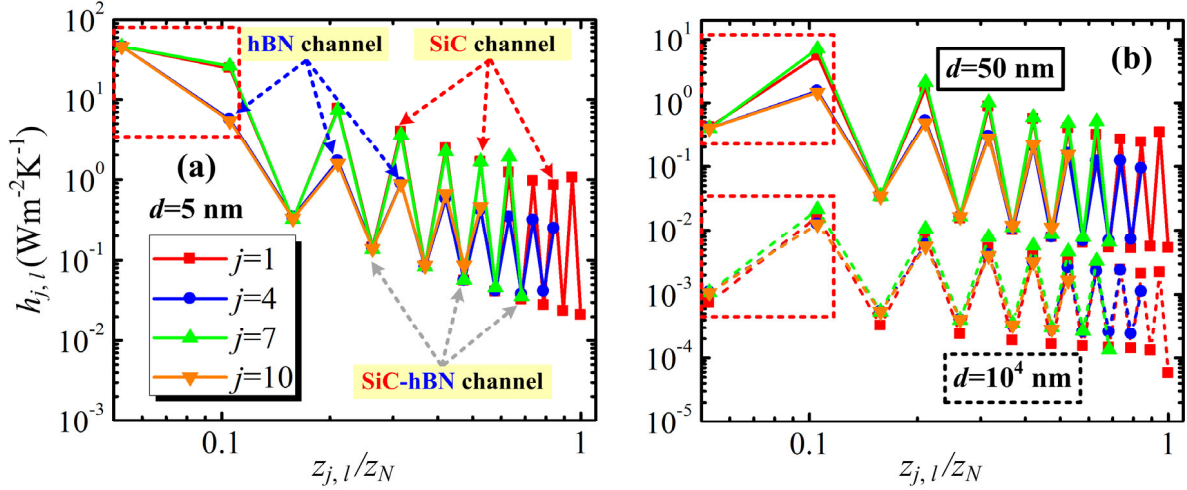

Fig. 4. Heat transfer coefficients between body $j$ ($j$ = 1, 4, 7, 10) and $l$ ($l > j$), where $j$=1 and 7 indicate SiC slabs and $j$=4 and 10 correspond to hBN slabs. (a) $d$ =5 nm; (b) $d$ = 50 nm and $10^4$ nm.

We now focus on the underlying mechanism causing the alternating pattern and the splitting in the temperature distribution. To avoid the explicit effect of the equilibrium temperatures in the analysis, we consider the heat transfer coefficients (HTCs) $h_{j,l}$ which are defined by

$$h_{j,l} = \frac{\varphi_{j,l}}{\Delta T} = \frac{1}{4\pi^2} \int_0^\infty \hbar\omega \frac{\partial n}{\partial T} d\omega \int_0^\infty \kappa \xi(\omega,\kappa)^{j,l} d\kappa \qquad (11)$$

in the limit of a vanishing $\Delta T$ around $T$ = 300 K, where $\varphi_{j,l}$ denotes the energy exchanged between bodies $j$ and $l$. In Fig. 4, we evaluate the HTC between bodies $j$ ($j$ = 1, 4, 7, 10) and $l$ ($l > j$), for which $j$=1 and 7 indicate SiC slabs, while $j$=4 and 10 correspond to hBN slabs. The horizontal axis in the figure represents the value of the ratio $z_{j,l}/z_N$, where $z_{j,l}$ is the distance between bodies $j$ and $l$. The first point from the left in each curve denotes the HTC between body $j$ and the adjacent body to its right, whereas the last point of each curve (on the right) represents the HTC between bodies $j$ and $N$. In the figure, we have explicitly indicated the HTCs between the hBN slabs, SiC slabs, and between SiC and hBN slabs, which are referred to as the hBN channel, SiC channel, and SiC-hBN channel, respectively. For $d$ = 5 nm, as shown in Fig. 4(a), the HTC between adjacent bodies (the first point from the left) is larger than the remaining coefficients for all three channels, so nearest-neighbor heat transfer is more important than the heat exchange with distant bodies for very short separation distances. The dashed-line squares in red in this figure highlights that the HTC between bodies $j$ and $j$+1 is larger than the HTC between bodies $j$ and $j$+2. According to this, the temperature distribution is dominated by the interaction with first neighbors and a monotonic behavior is obtained, as we have already shown in Fig. 2(c). Furthermore, in Fig. 4(b), we show the results of the HTCs for larger separations, taking $d$ = 50 nm and $10^4$ nm in both near-field and far-field regime. As emphasized by the dashed-line squares in red in this figure, the HTC between bodies $j$ and $j+1$ now is smaller than the HTC between bodies $j$ and $j+2$. It means that for separations beyond the extreme near field, the long-range coupling between distant objects made of the same material leads to a heat transfer which is more important than the heat transfer between closest neighbors made of different materials. As is clear from the plots, the SiC-hBN channel is subdominant and the HTCs reveal that both the hBN and SiC channels drive the heat exchange in the system, which happens for distances $d$ > 5 nm in our examples. Interestingly, the radiative heat exchange through these two channels induces two distinct components in the spatial distribution of temperatures that link bodies of the same material. This dual-channel photon heat exchange is responsible for the splitting of the temperature profiles we observe in Fig. 2(c).

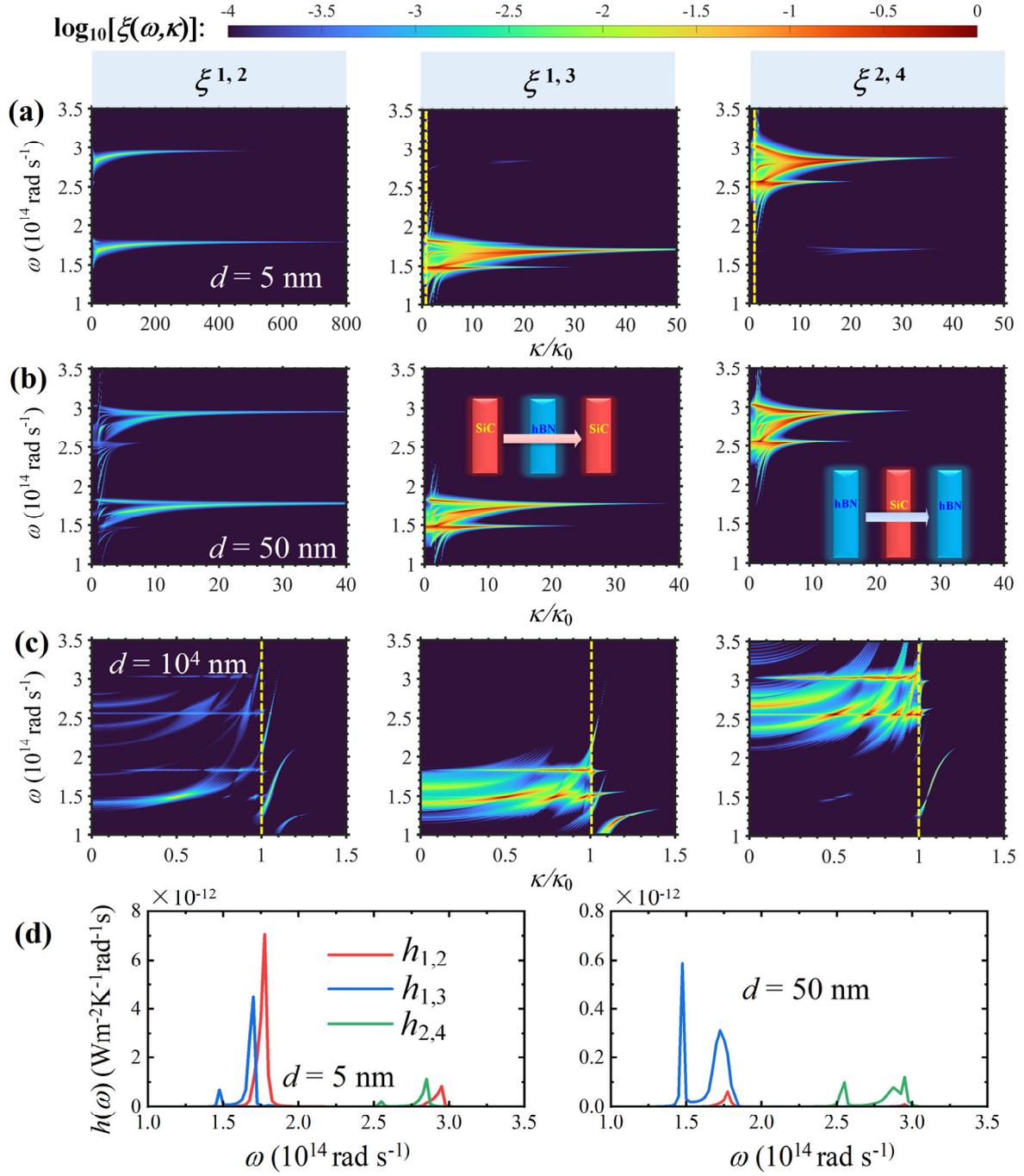

Fig. 5. The energy transmission coefficients at $d = 5$ nm (a), $d = 50$ nm (b), $d = 10^4$ nm (c) between different bodies: bodies 1 (SiC) and 2 (hBN) in the first column; bodies 1 (SiC) and 3 (SiC) in the second column; bodies 2 (hBN) and 4 (hBN) in the third column. Here $\kappa_0 = \omega/c$, $c$ being the speed of light. In (d) we show the spectral heat transfer coefficients for $d = 5$ nm and $d = 50$ nm.

To gain a comprehensive understanding of the behavior of the HTCs and the associated channels for the heat exchange, we analyze the energy transmission coefficients which describes the intensity of electromagnetic modes contributing the radiative heat transfer in the structure. These coefficients are shown in Fig. 5 represented in the frequency-wave vector space. We show the coefficient $\xi^{1,2}$ which is representative of the SiC-hBN channel and the coefficients $\xi^{1,3}$ and $\xi^{2,4}$ which account for the SiC and hBN channels, respectively, responsible for the splitting of the temperature distributions. The coefficients $\xi^{1,2}$, $\xi^{1,3}$, and $\xi^{2,4}$ are presented in the first, second, and third column of Fig. 5, respectively, for different separation distances. Let us first focus on the case $d = 5$ nm in Figs. 5(a). As seen in this figure for $\xi^{1,2}$, two resonances are observed which correspond to the surface phonon polaritons supported by SiC, the resonance at the lower frequency, and supported by hBN at the higher frequency. In the coefficients $\xi^{1,3}$ and $\xi^{2,4}$, only the resonance associated with the corresponding channel can be appreciated, while the other one is almost completely attenuated. From the plot of $\xi^{1,2}$ for $d = 5$ nm, we see that large values of the wave vector contribute to the heat transfer in the SiC-hBN channel, leading to peaks in the spectral heat transfer coefficients shown in Fig. 5(d), in such a way that this channel is stronger than the SiC and hBN channels quantified by $\xi^{1,3}$ and $\xi^{2,4}$, respectively. This causes that for $d = 5$ nm, as mentioned before, nearest-neighbor interactions dominate the heat transfer and therefore, a monotonic temperature distribution is obtained. In addition, the radiative heat transfer can be enhanced around resonance of SPhPs for thin films due to surface wave coupling [48-52]. The situation is different for larger separations, as seen in Figs. 5(b) and 5(c). For $d = 50$ nm in Fig. 5(b), we see that the SPhPs resonances for the SiC and hBN channels are more intense than the resonances excited in the SiC-hBN channel, the latter being damped by the mismatch in the frequencies of the SPhPs. This is corroborated in the plots of the spectral heat transfer coefficients in Fig. 5(d). For $d = 50$ nm, in the near field, the long-range dual-channel heat exchange controlling the splitting of the temperature distribution is realized through photon tunneling, as the heat transfer is mediated by the SPhPs. The mechanism behind the dual-channel heat exchange is different in the case $d = 10^4$ nm, for which the transmission coefficients are shown in Fig. 5(c), since the SPhPs are not excited for this separation. Instead, the heat transfer proceeds through propagating modes with frequencies mainly around the Reststrahlen band of each material.

## C. THERMAL RELAXATION

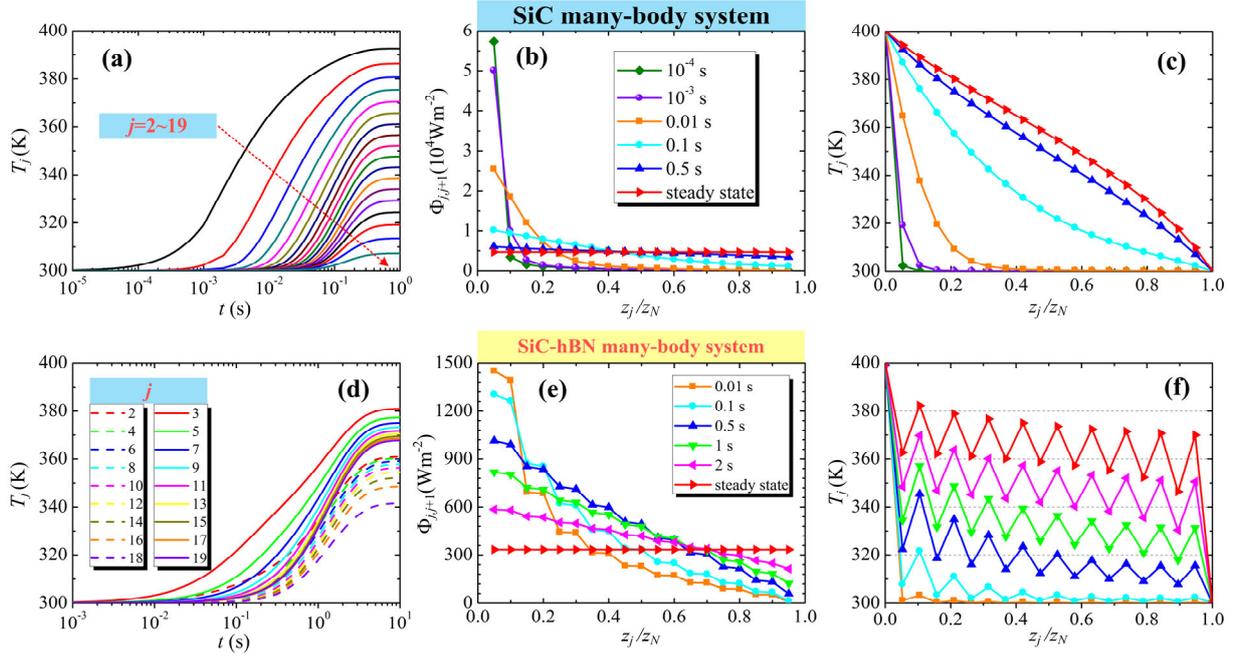

Fig. 6. Thermal relaxation dynamics for the SiC many-body system (first row) and the SiC-hBN many-body system (second row). In (a) and (d), we show the transient temperatures varying in time. In (b) and (e), the net radiative flux through the gap between slabs $j$ and $j+1$ is plotted for several times. In (c) and (f), the transient temperature distributions are shown at different moments.

In the present section, we investigate the thermal relaxation dynamics of the system. The temperatures of the bodies are collected in the vector $\boldsymbol{T}(t)$ at time $t$ and can be calculated by solving the energy balance equation

$$\partial_t \boldsymbol{T} = \mathbb{K} \cdot \boldsymbol{T} + \boldsymbol{S} \quad (12)$$

where $\mathbb{K} = \mathbb{H}/(C\delta)$ is a stiffness matrix defined in terms of the heat-transfer matrix $\mathbb{H}$, with $[\mathbb{H}]_{l,j} = h_{l,j}\,(l,j=1,...,N)$. Here $C$ = 8.15 Jcm$^{-3}$K$^{-1}$ and 1.846 Jcm$^{-3}$K$^{-1}$ are the heat capacity per unit volume of SiC and hBN, respectively [53, 54], and $\boldsymbol{S} = \dfrac{T_B}{C\delta}\left(h_{0,1} + h_{N+1,1},...,h_{0,N} + h_{N+1,N}\right)$ denotes the source term corresponding to the power supplied by the bath to each layer, where $T_B$ = 300 K is the ambient temperature. We assume that at the initial moment, all the slabs in the system have the same temperature $T$ = 300 K, and then the external slab on the left side of the system is heated so that its temperature is raised to $T_1$ = 400 K immediately. The results for the temperature dynamics are given for the many-body system composed of SiC slabs and for a system with SiC-hBN alternating slabs, respectively shown in the first and second rows in Fig. 6. The thermal relaxation in the system with hBN slabs is similar to that of SiC, so it is omitted here.

The transient temperatures are given in Figs. 6(a) and (d) for different bodies. We observe that the SiC-hBN many-body system needs more time to reach the steady state. To get insight into the heat transfer in the system from the initial moment to the steady state, we plot the net radiative flux transported through the gap between slabs $j$ and $j+1$, i.e., $\Phi_{j,\,j+1}$ in Figs. 6(b) and (e). The net radiative flux is higher near the left side of the system at the initial times, because of the sudden increase in the temperature of the first body, and gradually decays. In the meantime, this flux gets gradually larger at other positions and becomes uniform in a long-time limit. An interesting phenomenon can be observed in Fig. 6(e), in which $\Phi_{j,\,j+1}$ shows a step shape, meaning that the net heat flux decays with the position inside the system, just like walking down the stairs. The presence of thermal resistance distributed through the SiC-hBN system can be viewed as evidence of the dual-channel heat transfer regime discussed in the present work. Moreover, in Figs. 6(c) and (f), the transient temperature distributions are given at different times. As seen from the plots, the splitting of the temperature distribution develops from the initial moment to the steady state, as the temperature profiles always exhibit the alternating pattern. Thus, the effect is also present in transient states. The slow thermal relaxation observed in Fig. 6(d) together with the transient alternating pattern observed in Fig. 6(f) could be exploited to control thermal energy transport at the nanoscale.

## IV. CONCLUSION

In this work, we have demonstrated the splitting of temperature distributions in a many-body system composed of alternating slabs of different materials. The overall temperature distribution exhibits an alternating pattern that arises from the combination of two distinct components. Our analysis of heat transfer coefficients between bodies at different positions in the system reveals the existence of two different channels for heat transfer, which induce the splitting of the temperature distribution. These two channels stem from a long-range coupling between objects of the same material that allows for energy transport between distant parts of the system. We have also shown that in extreme near field these two components of the temperature distribution merge in a single distribution as the bodies tend to thermalize, thus leading to a monotonic behavior. Additionally, we have investigated the thermal relaxation dynamics of the system, revealing that the alternating pattern in the temperature distribution can also be realized in transient states. Our proposed dual-channel mechanism for controlling temperature distributions has potential applications in designing structures with specific thermal properties and developing new strategies for nanoscale thermal management.

## Declaration of competing interest

The authors declare that they have no known competing financial interests or personal relationships that could have appeared to influence the work reported in this paper.

## Data availability

Data will be made available on request.

## Acknowledgements

The supports of this work by the National Natural Science Foundation of China (No. 52206082), China Postdoctoral Science Foundation (No. 2021TQ0086), the Natural Science Foundation of Heilongjiang Province (No. LH2022E063), Postdoctoral Science Foundation of Heilongjiang Province (No. LBH-Z21013) are gratefully acknowledged.